\documentclass[aps,twocolumn,pra,floatfix,longbibliography]{revtex4-1}
\usepackage{amsmath}
\usepackage{amssymb}
\usepackage{bbm}
\usepackage[colorlinks=true,citecolor=blue, linkcolor=blue, urlcolor=blue]{hyperref}
\usepackage{graphicx,amssymb}
\usepackage{hyperref}
\usepackage[usenames, dvipsnames]{color}
\usepackage[dvipsnames]{xcolor}

\usepackage{soul}
\usepackage[normalem]{ulem}

\newcommand{\beq}{\begin{equation}}
\newcommand{\eeq}{\end{equation}}
\newcommand{\beqa}{\begin{eqnarray}}
\newcommand{\eeqa}{\end{eqnarray}}

\newcommand{\tx}[1]{\text{#1}}

\newcommand{\pd}{\partial}
\newcommand{\bmat}{\begin{matrix}}
\newcommand{\emat}{\end{matrix}}

\newcommand{\rar}{\rightarrow}

\newcommand{\mcal}[1]{\mathcal{#1}}
\newcommand{\floor}[1]{\left\lfloor{#1}\right\rfloor}

\begin{document}

\title{Layered Chaos in Mean-field and Quantum Many-body Dynamics}

\author{Marc Andrew Valdez$^1$}
\author{Gavriil Shchedrin$^1$}
\author{Fernando Sols$^2$}
\author{Lincoln D. Carr$^1$}

\affiliation{$^1$Department of Physics, Colorado School of Mines, Golden, CO 80401, USA \\ $^2$Departamento de Fisica de Materiales, Universidad Complutense de Madrid, E-28040, Madrid, Spain}

\date{\today}

\begin{abstract}
 {We investigate the dimension of the phase space attractor of a quantum chaotic many-body ratchet in the mean-field limit. Specifically, we explore a driven Bose-Einstein condensate in three distinct dynamical regimes - Rabi oscillations, chaos, and self-trapping  {regimes -}  and for each of them we calculate the correlation dimension. For the
ground state of the ratchet formed by a system of field-free non-interacting particles, we find four distinct pockets of chaotic dynamics throughout these regimes. We show that a measurement of local density in each of the dynamical regimes has an attractor characterized  {by} a higher fractal dimension, $D_{R}=2.59\pm0.01$, $D_{C}=3.93\pm0.04$, and $D_{S}=3.05\pm0.05$, compared to the global measure of current, $D_{R}=2.07\pm0.02$, $D_{C}=2.96\pm0.05$, and $D_{S}=2.30\pm0.02$. We find that the many-body case converges to  {the} mean-field limit with strong sub-unity power laws in particle number, $N^{\alpha}$, with $\alpha_{R}={0.28\pm0.01}$, $\alpha_{C}={0.34\pm0.067}$ and $\alpha_{S}={0.90\pm0.24}$ for each of the dynamical  {regimes} mentioned above. The deviation between local and global  {measurements} of the attractor's dimension corresponds to an increase towards  {higher} condensate depletion, which remains constant for long time scales in both Rabi and chaotic regimes. The depletion is found to scale polynomially with particle number $N$, namely  {as} $N^{\beta}$ with $\beta_{R}={0.51\pm0.004}$ and $\beta_{C}={0.18\pm0.004}$ for the two regimes.  {Thus, we find a strong deviation from the mean-field results, especially in the chaotic regime of the quantum ratchet.} The ratchet also reveals quantum revivals in the Rabi and  {self-trapping} regimes but  {not} in the chaotic regime, with revival times scaling linearly in particle number for Rabi dynamics. Based on the obtained results, we outline pathways for the identification and characterization of emergent phenomena in driven many-body systems. This includes
 {the} identification of many-body localization from the many-body measures of the system, the influence of entanglement on the rate of the convergence to the mean-field limit, and the establishment of  {a polynomial scaling} of the Ehrenfest time at which  {the} mean-field description fails to describe  {the} dynamics of the system.
}
\end{abstract}

\maketitle

\section{Introduction}\label{intro}

\par In recent years, periodically driven quantum systems have been the subject of extensive theoretical~\cite{kitagawa2010, lindner2011, ponte2015, abanin2015} and experimental~\cite{lignier2007, eckardt2009, zenesini2009, rechtsman2013} efforts. These systems allow the exploration of unique physical phenomena such as topological states of matter~\cite{kitagawa2010, lindner2011, rechtsman2013}, the ability to precisely tune quantum phase transitions~\cite{lignier2007, zenesini2009}, and localization~\cite{eckardt2009, ponte2015, bukov2015}. In close relation to localization, periodically driven quantum systems have also been instrumental in probing the understanding of various aspects of quantum chaos~\cite{moore1995, chabe2008, behinaein2006}. In particular, these dynamical systems have established firm connections between quantum signatures of chaos beyond Random Matrix theory by observing quantum dynamics that are comparable to phase space attractors of classical particle--like or semi-classical wave--like chaos~\cite{behinaein2006, kantz2004}. However, the role of particle interaction in many-body systems and their semi--classical counterparts has not been fully understood in this context. Such driven and interacting systems typically possess many relevant length and time scales that may give rise to a new realm of physical phenomena. In this article, we explore the dynamics of a periodically driven quantum many-body system in the mean-field limit for both regular and chaotic dynamics. We find that chaos in a quantum ratchet manifests  {itself} in a quantifiably different way depending on the length scale observed within the system, i.e., local measurements reveal higher dimensional attractors  {as} compared to global measurements.

\begin{figure}[h!]
\begin{center}
\includegraphics[width=\linewidth]{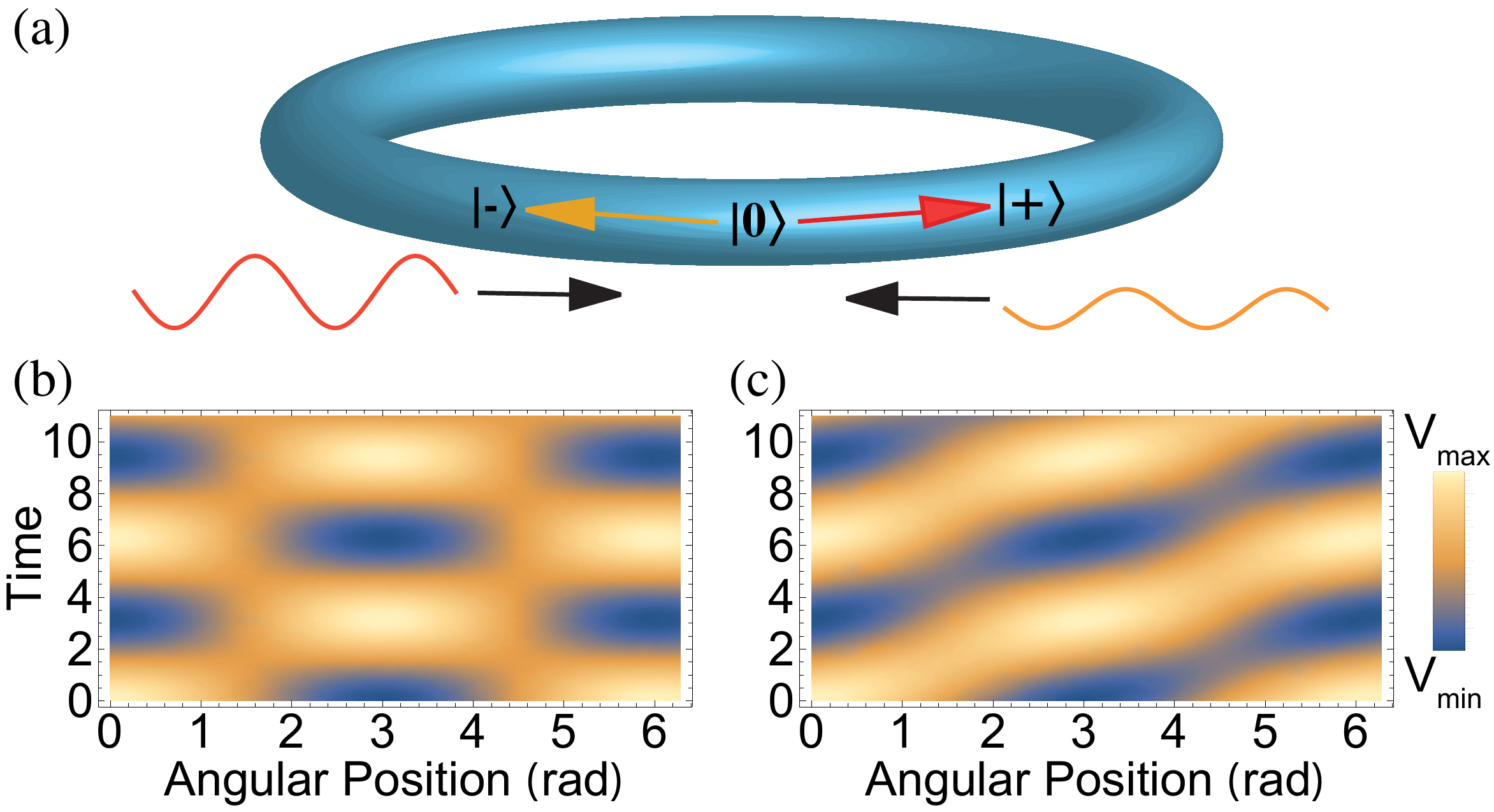}
\end{center}
\caption{\label{schem2} \emph{Quantum Ratchet via Toroidal Condensate} (a) Off center rotation of a toroidal Bose-Einstein condensate~\cite{henderson2009, ramanathan2011, moulder2012, ryu2013, ryu2014} generates the effect of driving by two counter propagating waves in one spatial dimension~\cite{heimsoth2013}. Only when the amplitude of the two waves is unequal can such a system present a ratchet effect due to asymmetric coupling to positive and negative angular momentum modes. (b) Space-time plot of the drive given in (a) with equal amplitude fields. (c) Once the drive amplitudes differ, the BEC becomes coupled to positive and negative modes, easy visible in the symmetries of the space-time plot for this case.}
\end{figure}

\par Specifically, we study an interacting many-body quantum ratchet that is known to be chaotic for a particular range of atomic interaction strengths and external driving parameters. This system consists of a Bose-Einstein condensate (BEC) under periodic driving that breaks generalized parity and time-reversal symmetries~\cite{heimsoth2012,heimsoth2013,valdez2017}, see Fig.~\ref{schem2}. The ratchet effect, or directionally biased motion, is induced by these symmetry violations~\cite{flach2000, reimann2002, denisov2014} and manifests in the particle current~\cite{heimsoth2012, heimsoth2013}. For near resonant driving this results in two regular dynamical regimes, Rabi oscillations for weak particle interaction and self-trapping for strong interactions, with a chaotic regime for intermediate couplings~\cite{heimsoth2012,heimsoth2013,valdez2017}. This system can be treated in the many-body framework of the Bose-Hubbard model with a time-dependent potential, or,  {alternatively}, in a time-independent truncated Floquet model~\cite{heimsoth2012, heimsoth2013, valdez2017}. In  {either} case, a well defined mean-field limit gives rise to nonlinear dynamics. Using the delay embedding method for calculating attractor correlation dimension~\cite{grassberger1983, fraser1986, theiler1990, sauer1991, kantz2004}, we investigate the phase space structures of the many-body models as well as their mean-field counterparts. We show agreement between the attractor dimension  {for} both  {mean-field} models. Moreover, we identify four distinct regions of interaction strengths that produce chaotic dynamics, contrasting  {its} identification in the spectral statistics and quantum many-body dynamics~\cite{valdez2017}.  {We shall point out, however, that the performed
analysis of the dimension of the strange attactor and the correlation dimension of the system  {are} meaningful only if the system is characterized with a nonzero dissipation. Indeed, the existence of strange attractors in the quantum ratchet indicates  {the} existence of an underlying structure that the system limits to in the presence of dissipation. We shall point out that the identification of strange attractors allows one to determine dynamical outcomes of the system~\cite{stockmann2000quantum}.}

\par Mean-field methods have been instrumental in the description of a wide range of many-body phenomena such as the  {Ginzburg--Landau} theory of superconductivity 
 {and the} Bogoliubov theory of superfluidity, among many others~\cite{trias2000, binder2000, alcala2017}.  {The success of the mean-field description of a wide spectrum of physical systems heavily relies on the fast convergence of  {the dynamics of finite sized many-body systems} with the number of particles that form the system}.  {However, whenever these systems exhibit chaotic dynamics, one usually considers either classical aspects of the system with the corresponding mean-field description, or the full quantum many-body treatment.} In the studies that have merged both of these descriptions, the physical systems have not been dominated by the interactions, and  therefore classical measures and parameters have been utilized in describing  {the} dynamics of the system~\cite{moore1995, chabe2008, behinaein2006}. In contrast, we take both approaches for the quantum ratchet system where interactions play the dominant role in defining the type of dynamics exhibited by the system. This allows us to differentiate in a quantitative way between different models that are used in the description of current experiments~\cite{ramanathan2011, moulder2012, ryu2014}.
Such differences can be seen even in regular dynamical regimes via large condensate depletion and subsequent quantum revivals. This situation only becomes worse for chaotic regimes, where the condensate depletes, and thus results in  {an} absence of quantum revivals.
 {The non-equilibrium dynamics of the condensate held at a finite temperature
and driven by external fields results in both thermal and dynamical depletion of the condensate. A number of powerful methods have been developed  {that allow one} to investigate thermal depletion of the condensate, including  {the} Hartree-Fock-Bogoliubov-Popov method \cite{hutchinson1997finite,dodd1998collective,giorgini1998damping}  {and the} Zaremba-Nikuni-Griffin method \cite{zaremba1999dynamics, griffin2009bose}, along with methods based on  {the} projected Gross-Pitaevskii \cite{davis2001simulations, simula2006thermal} and stochastic Gross-Pitaevskii equations \cite{weiler2008spontaneous, fialko2012quantum}. The self-consistent treatment of the dynamical depletion of a driven condensate held at a finite temperature
can be obtained within the second-order number-conserving method \cite{morgan2003quantitative, gardiner2007number, billam2012coherence, billam2013second}. The latter method  {allows one} to obtain  {a} number-conserving description of a driven condensate, preserve the orthogonality of condensate and non-condensate parts of the BEC using the Penrose-Onsager criterion of the Bose--Einstein condensation \cite{penrose1956bose}, and allows for the interaction and transfer of particles within the entire BEC. Specifically, the application of this method to a $\delta$-kicked-rotor condensate revealed that the unbounded growth of the non-condensate part of the condensate is a direct consequence of the instabilities in the linearized Gross-Pitaevskii equation. By incorporating the second-order corrections, this method enables
a self-consistent treatment of the condensate's back-action and shows that the growth of the non-condensate part is damped out, thus revealing the role of the dynamical depletion in a driven BEC \cite{gardiner2007number,billam2012coherence, billam2013second}.
}

\par Advances in the experimental control of ultra-cold atoms have opened up a wide variety of new research directions~\cite{bloch2008}. In particular, studies have demonstrated novel phenomena including persistent currents~\cite{ramanathan2011, moulder2012, ryu2014}, precision and dynamic control of optical traps~\cite{henderson2009, ryu2013, ryu2014}, the generation of quantum ratchets~\cite{mennerat1999,salger2009}, and  {the} measurement of quasi-classical phase space structures~\cite{behinaein2006}. Such techniques, along with the knowledge that mean-field models have been well tested and validated~\cite{trias2000, binder2000, alcala2017}, allow for new perspectives on the connection between linear quantum many-body dynamics and mean-field nonlinear dynamics in the out of equilibrium regimes. Furthermore, they allow investigations into dynamical features of quantum chaos beyond the well studied single particle case~\cite{moore1995, chabe2008, behinaein2006}.  {Here we provide a detailed study of the mean-field and quantum many-body dynamics and find that both treatments lead to the same sub-unity power  {law} time scaling.}  {Furthermore, we provide a context in which the deviation  {from the mean-field description} with the subsequent recovery of the mean-field phase space structures can be observed via long lived condensate depletion and quantum revivals.}

\section{Many-body and Mean-Field Models}

\par In this paper, we consider a BEC in a toroidal trap~\cite{henderson2009, ramanathan2011, moulder2012, ryu2013, ryu2014}, resonantly driven by a generalized parity and time-reversal symmetry breaking potential that is periodic in time. We will focus on a potential that can be experimentally realized by
 {rotating the entire condensate confined in a trap around a circle of radius $r$.}
~\cite{heimsoth2013, valdez2017}.  {This driving, together with repulsive particle interactions, produces three distinct dynamical regimes for a particle current: Rabi oscillations, quantum  {chaotic}, and self-trapping  {regimes}, in accordance with increasing interaction strength.} This system can be effectively treated using a one dimensional Bose-Hubbard model with periodic boundary conditions, which can be viewed as a discretization of the torus or explicit lattice sites on a ring~\cite{valdez2017}. The Hamiltonian for this model is
\begin{align}\label{BHH}
\hat{H}_\text{B}=&-J\sum_{j=1}^L( \hat{b}_{j}^{\dag}\hat{b}_{j+1}+\text{h.c.})+\frac{U}{2}\sum_{j=1}^L \hat{n}_j(\hat{n}_j-\hat{\bf1}) \nonumber \\
&+\sum_{j=1}^L V_j(t) \hat{n}_j
\end{align}
where the driving field, $V_{j}(t)$, is given by,
\begin{align}\label{drivepot}
V_j(t)&=E_+\cos(\kappa r \theta_j-\omega t)+E_-\cos(\kappa r \theta_j + \omega t).
\end{align}
Here $\hat{b}_j^{\dag}$, $\hat{b}_j$, and $\hat{n}_j=\hat{b}_j^\dag \hat{b}_j$ are the bosonic creation, annihilation, and number operators on the $j^{\text {th}}$ site, respectively. These operators obey the standard bosonic commutation relations $[\hat{b}_i,\hat{b}_j^\dag]=\delta_{ij}$ and $[\hat{b}_i,\hat{b}_j]=0$. Each site in the $L$ length discretization corresponds to a position $r \theta_j$, where $r$ is the radius of the torus, $\theta_j \in [0,2\pi)$ is the angle to the site, and the periodic boundary conditions of the torus impose $\theta_{j+L}=\theta_j$. The coefficients $E_\pm$ give the amplitude of the driving as well as  {control} the violation of P and T symmetries when they are not equal. $\kappa$ is the wave number of the driving field and $\omega$ is the frequency. For simplicity we consider resonant driving with the first harmonic of the toroidal trap, $\kappa = 1/r$, and $\omega = 2J[1-\cos(2\pi/L)]/\hbar$. A more detailed list of drive configurations were considered in~\cite{heimsoth2013}, while the specific choice of the potential in \eqref{drivepot} was made based on its simplicity and experimental relevance~\cite{heimsoth2013, valdez2017}.

 {
The experimental realization of the quantum ratchet
\cite{grossert2016experimental, salger2009, kenfack2008controlling, chien2015quantum, zhan2011quantum}
with the observation of its characteristic dynamical regimes can be achieved by a Bose-Einstein condensate of $^{87}\text{Rb}$ loaded in a ratchet potential given by Eq.~(\ref{drivepot}) formed by two counter-propagating laser fields. Upon loading the BEC into the optical potential, the atomic cloud undergoes a free expansion with a subsequent imaging of the BEC via the well-established time-of-flight (TOF) technique. The TOF-generated images of the atomic cloud reveal the atomic-velocity distribution. This allows one to obtain not only  {the} mean atomic momentum, but also  {the} time evolution of the particle current and  depletion of the condensate as a function of the coupling parameter $U(N-1)/L$. The periodic driving of the system and its dynamical response give rise to two time scales exhibited by the quantum ratchet. The first time scale is defined by the driving period, $T=2\pi/\omega$, while the non-interacting Rabi period, $T_\tx{R}=2\pi/(E_+^2 + E_-^2)^\frac{1}{2}$, defines the second time scale.  {The inclusion of interaction  {in} the system results in three distinct dynamical regimes: Rabi oscillation for weak interactions, the onset of self trapping for strong interaction $U(N-1) = 2 L J \tx{max}(E_+, E_-)$, and finally the chaotic regime that is exhibited by the system for the intermediate interaction strengths~\cite{heimsoth2013, valdez2017}.}}

\par It has been previously shown that the static and dynamic signatures of quantum chaos in our ratchet are preserved in a truncated three level system (3LS) defined by the
Floquet modes of the system~\cite{valdez2017, heimsoth2013}. This model can be derived by applying the $(t,t')$-formalism to the equation of motion for the second quantizated field operator $\hat{\psi}(x)$. This method takes the time-dependence induced by the periodic driving and absorbs it into an auxiliary parameter $t'$ in a way that makes the equation of motion take the form of a Floquet operator in $t'$. By expanding in Floquet states, a time-independent representation of the system is acquired~\cite{heimsoth2012,heimsoth2013}. In the special case of weak driving, one can restrict the expansion to the three lowest  {angular momentum} modes and obtain an effective three level model (3LS)~\cite{heimsoth2013}. In this picture the effective Hamiltonian is,
\begin{align}\label{3LH}
\hat{H}_{\text{3LS}}&=\frac{E_+}{2}( \hat{a}^{\dag}_+\hat{a}_0+\text{h.c.})+\frac{E_-}{2}( \hat{a}^{\dag}_-\hat{a}_0+\text{h.c.})\nonumber\\&-\frac{U}{2L} \sum_{\nu} \hat{n}_{\nu}(\hat{n}_{\nu}-\hat{\bf 1}),
\end{align}
where $\hat{a}_\nu^\dag$, $\hat{a}_\nu$, and $\hat{n}_\nu$ are bosonic creation, annihilation, and number operators, respectively, for the  {angular momentum} mode $\nu$, which satisfy the bosonic commutation relations, $[\hat{a}_\mu,\hat{a}_\nu^\dag]=\delta_{\mu\nu}$ and $[\hat{a}_\mu,\hat{a}_\nu]=0$. Here the index of the operator represents positive, negative, and zero  {angular momentum} modes. We note that interactions that are repulsive in the Bose-Hubbard model become attractive in the angular momentum representation of the Floquet modes~\cite{heimsoth2013} and the factor of $L$ comes from the $L$ length spatial discretization  {of} the Bose-Hubbard model.

\par Both of these models allow for a corresponding mean-field model in the limit where $N\rar \infty$ and $U\rar0$ with  {$U(N-1)=\tx{const}$}. These models can be acquired by first calculating the equation of the boson destruction operators, and, under the usual assumptions, taking the expectation value with respect to a tensor product of Glauber coherent states~\cite{mishmashthesis}. The result for the Bose-Hubbard model is the non-integrable discrete nonlinear Schr\"{o}dinger equation (DNLS)~\cite{mishmashthesis},
\beq
i\hbar \frac{\pd}{\pd t} \phi_j= -J(\phi_{j+1}+\phi_{j-1}) + V_j(t) \phi_j +  N U |\phi_j|^2 \phi_j,
\eeq
where $\phi_j$ is the coherent state amplitude at the $j^{\text {th}}$ site, with $\sum_j |\phi_j|^2=1$, and we are in the mean-field limit. For the 3LS model, we arrive at a similar DNLS type model with only three amplitudes which we will call the 3GP~\cite{heimsoth2013},
\beq
i\hbar \frac{\pd}{\pd t} \left[\bmat \phi_+ \\ \phi_0 \\ \phi_- \emat \right] = \left[\bmat \frac{-N U}{L}|\phi_+|^2 & \frac{E_+}{2} &  0 \\ \frac{E_+}{2} & \frac{-N U}{L}|\phi_0|^2 & \frac{E_-}{2} \\ 0 & \frac{E_-}{2} & \frac{-N U}{L}|\phi_-|^2 \emat \right]  \left[\bmat \phi_+ \\ \phi_0 \\ \phi_- \emat \right],
\eeq
where $\phi_\nu$ is the coherent state amplitude in the  {angular momentum} mode $\nu$, we are again in the mean-field limit, and $\sum_\nu |\phi_\nu|^2=1$.

 {In our previous work~\cite{valdez2017}, we have employed hopping units where time is scaled by $\hbar/J$ and energies are scaled by $J$. However, for the clarity of the present work we will not assume $\hbar=1$, and will use the parameters $E_+=0.0225 J$, $E_-=0.0075 J$ and vary the mean-field interaction $NU/J$ in the range $[0, 0.325]$, which  {covers} all three dynamical regimes.} For our simulations we use the ground state of the non-interacting system with no driving and time evolve for a duration of 1000$T_\tx{R}$. In the DNLS we use $L=6$ in order to capture an extended range of angular momentum modes, which we simulate using a fourth order Runge-Kutta with a sufficiently converged time step. The 3LS has a much smaller Hilbert space than the Bose-Hubbard, scaling with $(N+1)(N+2)/2$ compared to ${{N+L-1}\choose{L-1}}$, respectively. This allows for long time dynamics for $N$ up to 40 particles in the 3LS using exact diagonalization. For the 3GP we use a fourth order Runge-Kutta method with a sufficiently small time step,  {$\Delta{t}$}, in order to converge results.
 {We  {} emphasize that the convergence of the numerical solution for both 3GP and  DNLS directly corresponds to the truncation error of the fourth order Runge-Kutta method, $\epsilon_{\text{RK}}$, which scales as $\epsilon_{\text{RK}}=\mathcal{O}(\Delta{t}^{4})$ \cite{butcher2016numerical, iserles2009first}.}

\section{Nonlinear methods} \label{NLMethods}

\par Extensive research in nonlinear dynamics  {has} revealed a number of fundamental results that firmly established the numerical study of dynamical systems and chaos.  {The prime example among them is the Takens's theorem, which provides the necessary conditions under which the evolution of a dynamical system can be obtained from a
discrete time series of the state of a dynamical system~\cite{takens1981, sauer1991, theiler1990, kantz2004}.} This makes it possible to not only calculate attractor dimensions, but also determine Lyapunov exponents from experimental data~\cite{kantz2004}. Similar methods have also been developed that can identify the chaotic nature of a time series without having to calculate the full spectrum of Lyapunov exponents~\cite{gottwald2009valid, gottwald2009imp}. For our study, we will consider the ${0}-{1}$ test for identification of the chaotic regime~\cite{gottwald2009valid, gottwald2009imp}, and the delay embedding method for estimating the correlation dimension of an attractor~\cite{takens1981,sauer1991,theiler1990,kantz2004}.

\subsection{${0}-{1}$ Test For Chaos} \label{01Test}

\par  {In this section we will introduce the binary ${0}-{1}$ test for chaos identification in a dynamical system that yields zero for a regular dynamics and unity for a chaotic dynamics.} The underlying concept is to map a length $\mcal{N}$ time series $X=\{x_1,x_2, \cdots, x_\mcal{N}\}$ onto an effective two dimensional phase space with coordinates $(p,q)$ where a chaotic system's trajectory will manifest as the Brownian type motion with characteristic linear scaling of mean squared displacement with time~\cite{gottwald2009valid, gottwald2009imp}. This map is defined as
\beqa\label{coordinates1}
p_j &=& \sum_{i=1}^j x_i \cos(i c), \\
q_j &=& \sum_{i=1}^j x_i \sin(i c),
\eeqa
for $j\in \{1,2,\ldots,\mcal{N}\}$ where $c$ is an arbitrary parameter on the interval $(0, \pi)$ (see Fig~\ref{01coord}).  {From the mapped coordinates given by Eq.(\ref{coordinates1}) one then
proceeds with the evaluation of the modified mean squared displacement,}
\beq
M(j) = D(j) + W_\tx{osc}(j).
\eeq
Here, $D(j)$ is the normal mean squared displacement for points separated by $j$ time steps in the series of $p$ and $q$,
\beq
D(j) = \lim_{\mcal{N}\to\infty} \frac{1}{\mcal{N}} \sum_{i=1}^\mcal{N}
\left[
(p_{i+j}-p_i)^2+(q_{i+j}-q_i)^2
\right],
\eeq
and the counter-oscillatory function $W_\tx{osc}(j)$ is give by,
\beq
W_\tx{osc}(j)=\left(\lim_{\mcal{N}\to\infty} \frac{1}{\mcal{N}} \sum_{i=1}^\mcal{N} x_i\right)^2 \frac{1-\cos(j c)}{1-\cos(c)}.
\eeq
 {The counter-oscillatory function $W_\tx{osc}(j)$ removes the oscillatory dynamics of the mean squared displacement $D(j)$, while preserving the original asymptotic growth
of the modified mean squared displacement $M(j)$~\cite{gottwald2009imp}}. Since chaotic dynamics is mapped to Brownian-like motion, $M(j)$ scales linearly with $j$ approaching infinity,  {Thus,} one can introduce auxiliary vectors $\Lambda=(1,2,\ldots,\floor{\mcal{N}/10})$ and $\Delta=(M(1),M(2),\ldots,M(\floor{\mcal{N}/10}))$,  {and} the final binary ${0}-{1}$ test output acquires a particularly simple form,
\beq
K_c = \frac{\tx{cov}(\Lambda,\Delta)}{\sqrt{\tx{var}(\Lambda)\tx{var}(\Delta)}},
\eeq
where the covariance, $\tx{cov}(\Lambda,\Delta)$, between two vectors $\Lambda$ and $\Delta$ is defined as,
\beq
\tx{cov}(\Lambda,\Delta) \equiv \frac{1}{(\floor{\mcal{N}/10})}\sum_{j=1}^{\floor{\mcal{N}/10}} (\Lambda_j-\overline{\Lambda})(\Delta_j-\overline{\Delta}).
\eeq
 {Here $\overline{\tx{\bf x}}$ and   $\tx{var}(\tx{\bf x})=\tx{cov}(\tx{\bf x},\tx{\bf x})$ are the mean value and variance of the vector $\tx{\bf x}$, correspondingly.} We note here that by taking $j\in\{1,\cdots, \floor{\mcal{N}/10}\}$ the scheme preserves the limit structure of the mean squared displacement~\cite{gottwald2009valid, gottwald2009imp}. The test is then carried out for $K_c$ using a sample of $c$ values in the range $(0, \pi)$, which ensure the binary ${0}-{1}$ test is independent of the particular choice of $c$~\cite{gottwald2009valid}.  {We should mention that a discrete time series which is obtained from a dynamical system is characterized by an upper bound $c$ beyond which the power spectrum decays to zero for large frequencies $f$. This results in a maximal value of the upper bound, $c_\tx{max}=2\pi f_\tx{max}/f_\tx{s}$, which depends on the maximal frequency, $f_\tx{max}$, and the
the sample frequency, $f_\tx{s}$.} This limit is imposed in order to prevent false negatives of identifying chaos in the system, since every bound $c$ exceeding the maximal value,  $c>c_\tx{max}$, will result in periodic values for $p$ and $q$ and thus map phase space to a torus~\cite{gottwald2009imp}. It is also convenient to select an irrational number for $c$ such that it is not resonant with the frequencies of the time-series being tested.

\subsection{Correlation Dimension} \label{CDim}

\par The $0-1$ test allows us to determine the interaction strengths for which our system exhibits chaotic dynamics. In this section we will apply delay embedding techniques in order to determine the fractal dimension of the attractors for various measures of the quantum ratchet. There are multiple types of fractal dimensions, i.e., the box counting, information, and correlation dimensions~\cite{kantz2004}. All of these use the scaling of some measure on a dynamical system attractor with neighborhood size in phase space to calculate the dimension. Due to the computational accessibility, we will focus on the correlation dimension in order to characterize the dynamical regimes beyond the presence of chaos.  {In order to calculate the correlation dimension, $D_2$, we take a time-series $X$ and construct the delay embedding~\cite{theiler1990, kantz2004}.} This takes our time series and maps it to an $m$-dimensional phase space vector for each instant in time, that is $X\rar X'$ with
\beq
X' = \{v_1^\tau,v_2^\tau,\cdots,v_{\mcal{N}-(m-1)\tau}^\tau\}
\eeq
where $v_i^\tau=(x_1, x_{1+\tau}, \cdots, x_{1+(m-1)\tau})$, $\tau$ is the time delay, which must be selected so that the attractor is sufficiently unwrapped~\cite{theiler1990, kantz2004}. Using the delay vectors $v_j^\tau$ in $X'$ we can calculate the correlation sum~\cite{theiler1990, kantz2004},
\beq
C(m,\epsilon) = \frac{1}{P}\sum_{j=m}^P\sum_{k<j-w} \Theta(\epsilon - |v_j^\tau-v_k^\tau|),
\eeq
where $\epsilon$ is the radius of an $m$-dimensional  {sphere}, $P$ is the number of pairs of vectors used, $\Theta(\epsilon)$ is the Heaviside step function, and $w$ is the Theiler window, which ensures that all of the points taken into account are sufficiently uncorrelated in time~\cite{theiler1990, kantz2004}. With the proper selection of parameters, large enough embedding dimension, and a sufficiently sampled attractor, the correlation sum scales polynomially with the radius of the $m$-dimensional  {sphere}, where the power is the correlation dimension, that is, $C(m,\epsilon)\propto \epsilon^{D_{2}}$~\cite{theiler1990, kantz2004}. Generally, one does not have
the {\it a priori} value for $m$ to get a proper estimate for the correlation dimension $D_{2}$.
Thus, the convergence of  {the} $m$ value serves as an indicator for the correct estimate of the correlation dimension $D_{2}$~\cite{theiler1990, kantz2004}. A typical method to select the time delay $\tau$ is to take the first minimum of the mutual information of the time-series~\cite{fraser1986}, which will be used in our analysis. For the Theiler window we select $w=\mcal{N}^{1/2}\approx 30T_\tx{R}$, where $T_\tx{R}$ is the Rabi period,
 {which ensures that the temporal correlations do not produce spurious attractor dimensions}. We also check each parameter against perturbations in order to confirm that the attractor is invariant under smooth transformations~\cite{anderson2014}.

\section{Layered Chaos in a driven quantum ratchet}

\par We begin our study by characterizing the mean-field dynamics of our quantum ratchet in the 3GP and the DNLS. Using the ${0}-{1}$ test we first characterize the interaction ranges for which chaotic dynamics is present. We then move to calculate the fractal dimension of the attractors throughout the dynamical regimes of our system. For each method, we use a time-series of length $\mcal{N}_I=10^5$ for particle current, and in the DNLS we use $\mcal{N}_N\approx10^6$ for the local density at the third site, $|\phi_3|^2$.

\begin{figure}[t!]
\begin{center}
\includegraphics[width=\linewidth]{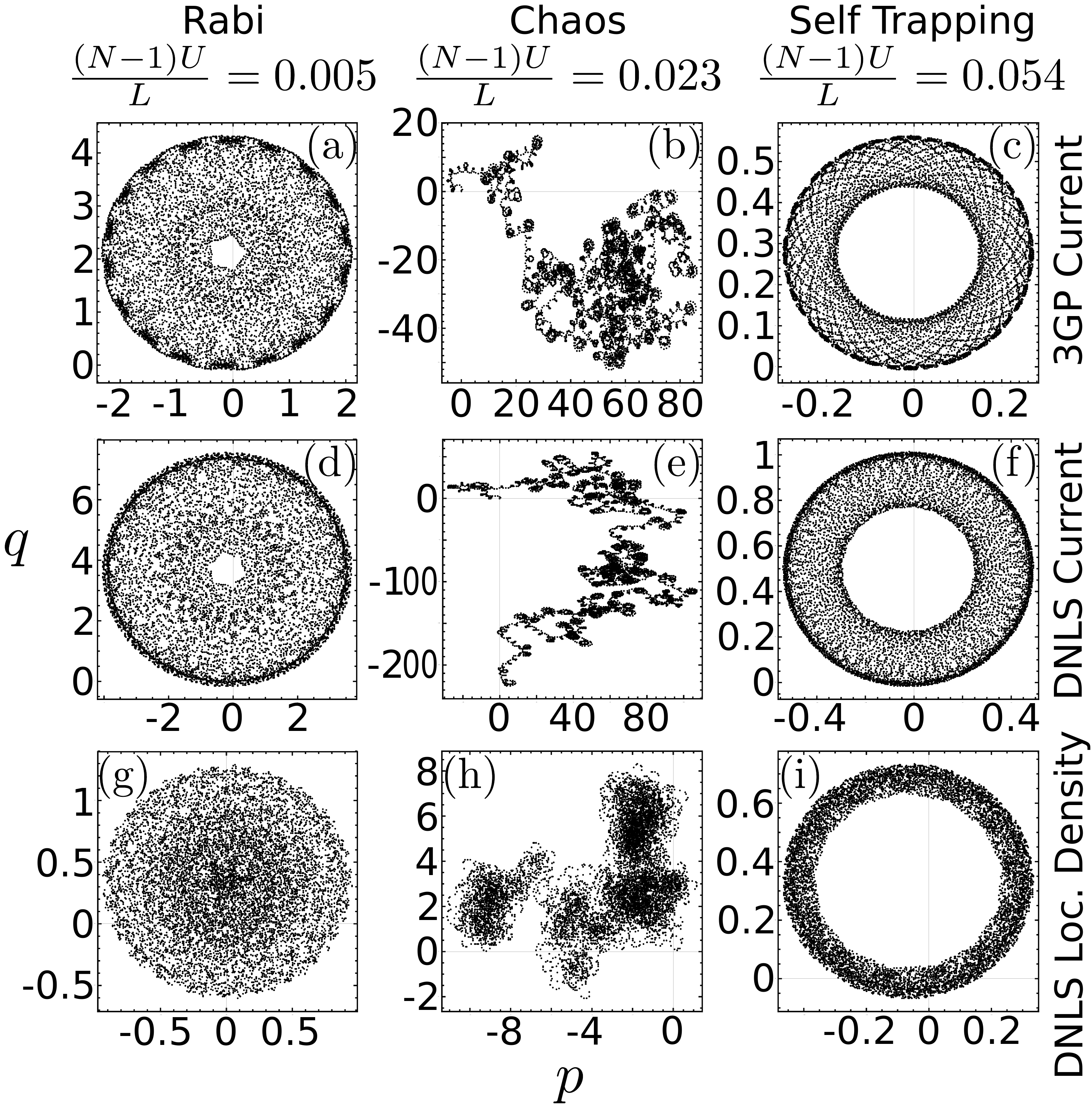}
\end{center}
\caption{\label{01coord}\emph{${0}-{1}$ Test Coordinates.} (a)-(i) Examples of the mapped coordinates in the ${0}-{1}$ test for chaos for the 3GP and six site DNLS for all dynamical regimes. As expected in the regular regimes, the Rabi (left) and self-trapping (right) dynamics are mapped to tori in the $(p.q)$ plane. The center column, corresponding to the chaotic regime, displays random walk like behavior with the points spreading over the effective phase space. This diffusive behavior is characteristic of chaotic dynamics under the map used in the ${0}-{1}$ test, making use of linear scaling in the mean squared displacement to identify chaos.}
\end{figure}

\par
\begin{figure}
\begin{center}
\includegraphics[width=\linewidth]{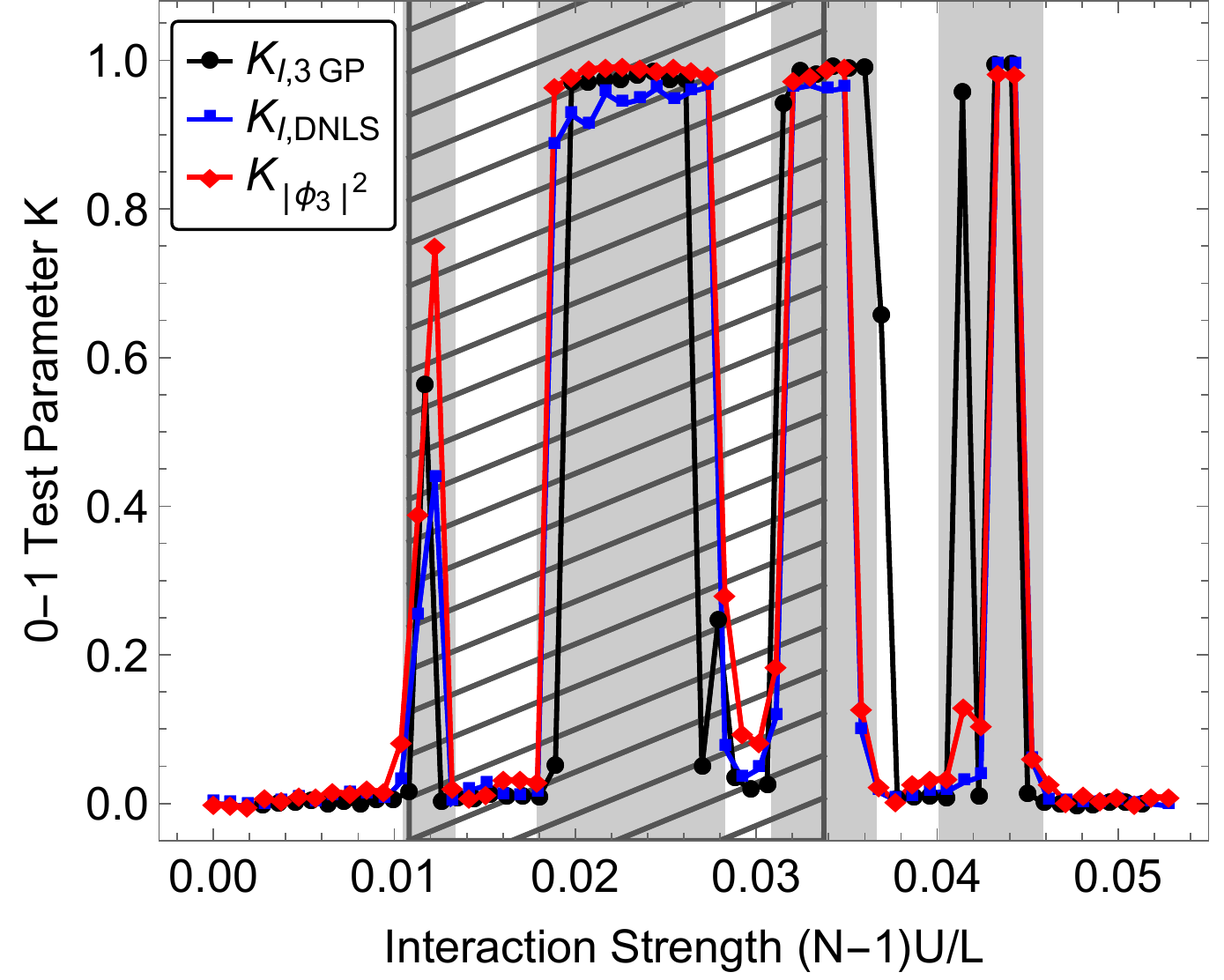}
\end{center}
\caption{\label{01test}\emph{${0}-{1}$ Test For Chaos.} The output of the ${0}-{1}$ test for chaos, $K$, shows distinct pockets of chaos (gray shading) that are the same for each model  {and} measure, with no chaos observed after the transition to self-trapping. This is in contrast to the quantum many-body dynamics and level statistics in~\cite{valdez2017}, which predicted one bulk region given by the gray hashed region. The current and local density in DNLS agree exceptionally well with the 3GP, indicating that  {the} inclusion of higher angular momentum modes in the DNLS does not affect the presence of chaos.}
\end{figure}

 In order to avoid resonances, we use a total of 100 $c$ values at multiples of the golden ratio, $\varphi=(1+\sqrt{5})/2$, for calculating the test parameter $K_c$, from which the median is taken as the final test value. For the maximum $c$ value imposed by the ${0}-{1}$ test for chaos, we have set $f_\tx{max}=2 \varphi \Omega_\tx{R}\approx 3.24\Omega_\tx{R}$, with $\Omega_\tx{R}$ the Rabi frequency, since  {for} frequencies  {higher than the} Rabi frequency the power spectrum  {decays} to zero. In Fig.~\ref{01coord}  {we provide} examples of the effective phase space coordinates $p$ and $q$ for measurements of current in both of our mean-field models and local density in the DNLS over each dynamical regime with $c=0.8 \pi/100 \approx 0.025$. In Fig.~\ref{01test} we give the final test parameter. We note that there are distinct pockets of chaos for the particle current in both mean-field models as well as the local density of the DNLS. This is contrary to the single dominant feature the level statistics and dynamical measures display in~\cite{valdez2017}, which is indicated by the gray shaded region in Fig.~\ref{01test}.  {However, the two models are in agreement for the regions which are chaotic, even though they arise from quantum models with quite different assumptions and completely different sets of approximations.}

\begin{figure}
\begin{center}
\includegraphics[width=\linewidth]{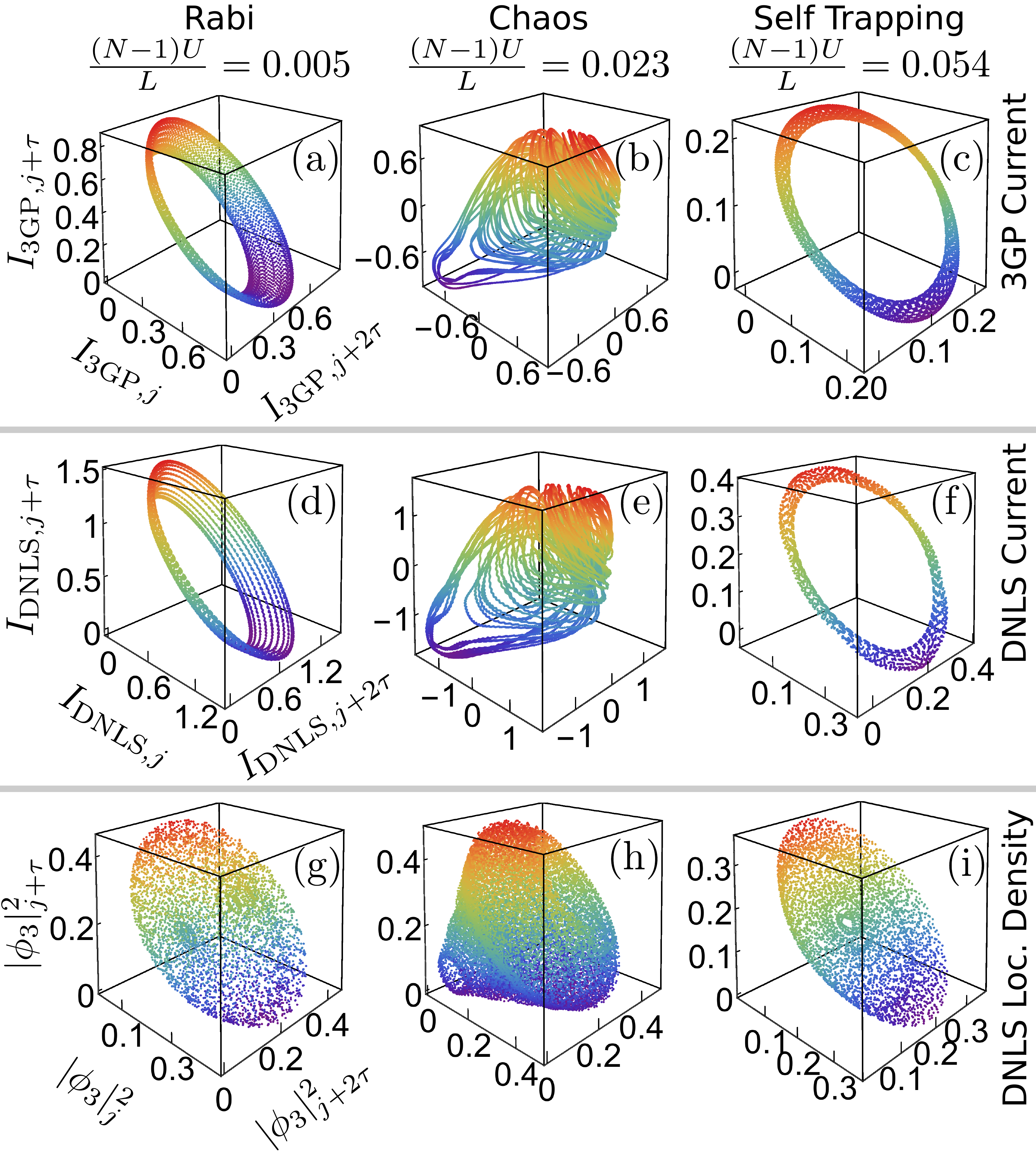}
\end{center}
\caption{\label{attractors}\emph{Delay Reconstructed Attractors.} (a)-(c) and (d)-(f), the current in both the 3GP and DNLS, respectively, have qualitatively similar attractors for each dynamical regime. The differences in coordinate ranges are due to the exclusion of the scaling factors for the angular momentum modes in the 3GP, which has no effect on the attractor dimension. (g)-(i) The local density in the DNLS has attractors that are visibly higher dimensional when compared to the current. In the Rabi and self-trapping regimes we see densely filled surfaces with no holes, unlike the  {currents}, which appear as tori. The attractors for the chaotic regime all show signs of self similarity, and have deviated from regular torus shapes to a higher dimensional region in phase space. Coloring is meant only to provide contrast to aid the eye. }
\end{figure}

\par
\begin{figure}
\begin{center}
\includegraphics[width=\linewidth]{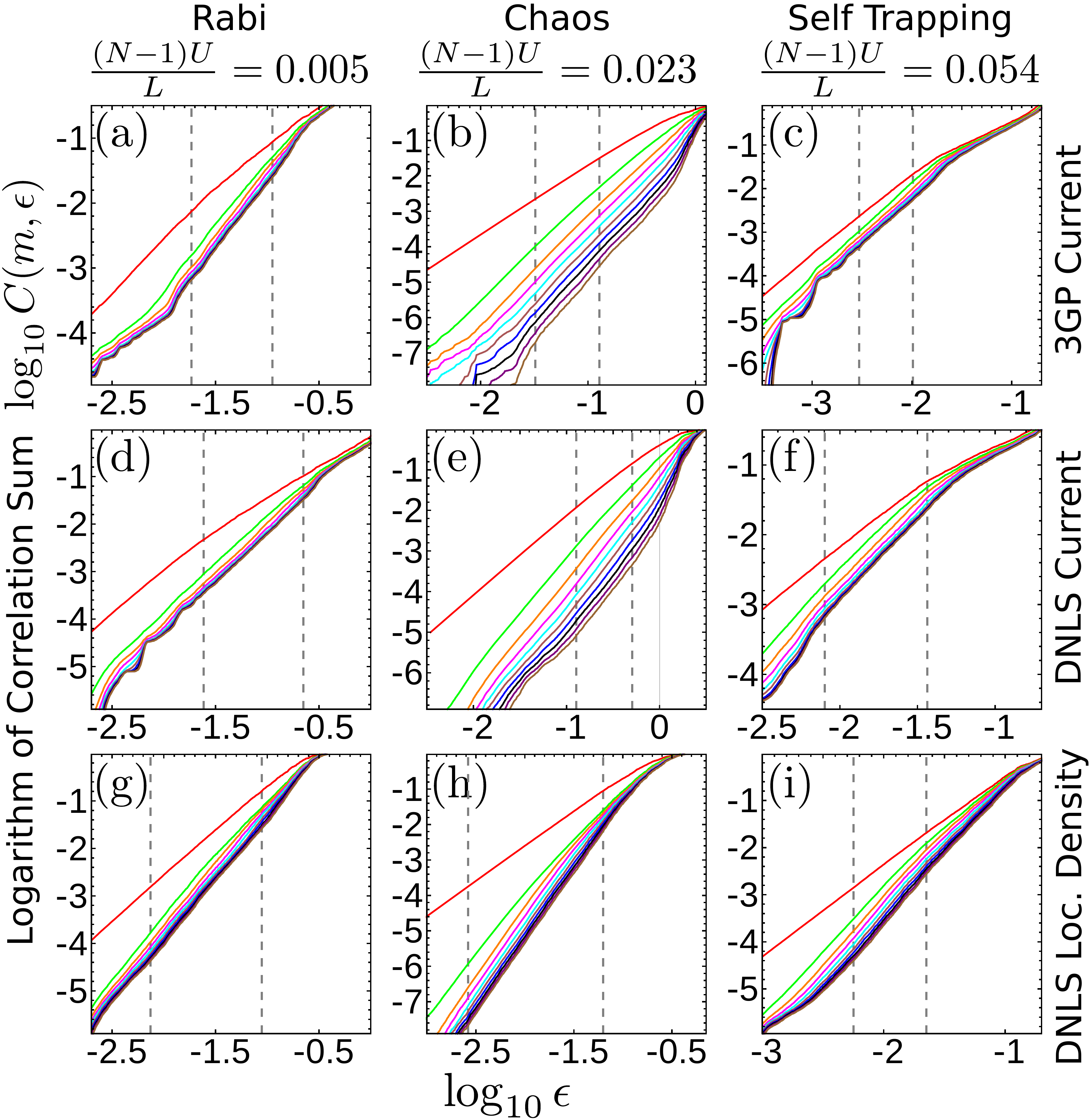}
\end{center}
\caption{\label{corrsum}\emph{Correlation Sums.} (a)-(i) Gray dashed lines indicate the start and end of the linear scaling regions used to calculate the correlation dimension, with the embedding dimension decreasing down and to the right. Each model and measure approaches a constant slope in the indicated region, with the range of $\epsilon$ checked for being within the size of delay reconstructed attractors.}
\end{figure}

\begin{figure}
\begin{center}
\includegraphics[width=\linewidth]{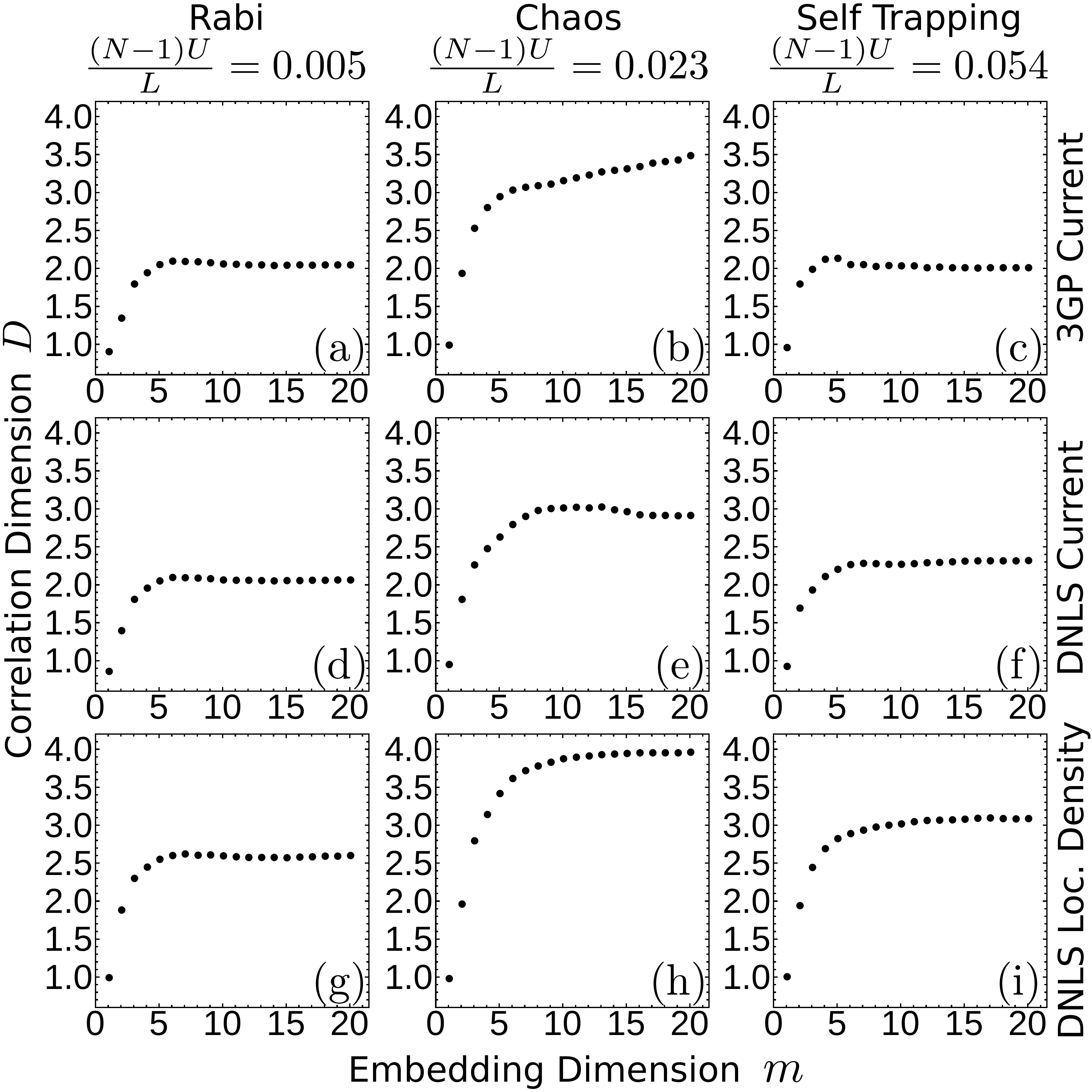}
\end{center}
\caption{\label{d2plat}\emph{Converging Correlation Dimension.} Each panel corresponds to the slope of the correlation sum given in Fig.~\ref{corrsum} with the same letter. Each model and interaction strength shows a plateau corresponding to the convergence of the correlation dimension in  {the} embedding dimension. The final correlation dimension is given by the average over all $m$ after the plateau is reached, see Tab.~\ref{d2ex} for numerical values. The slow increase in dimension as $m$ increases after the plateau, as seen in (b) and (i) is a typical artifact of noise in the time-series~\cite{theiler1990}.}
\end{figure}

 We now calculate the attractor dimension for the quantum ratchet using the correlation method laid out in Sec. ~\ref{CDim}. In Figures~\ref{attractors}-\ref{d2plat}, each sub-panel corresponds to the same time-series as seen in Fig.~\ref{01coord}, respectively, giving an example of all three regimes in both models at different scales of measurement for the DNLS. Each delay reconstruction uses the first minimum of the mutual information to calculate the time delay $\tau$~\cite{fraser1986}. In Fig.~\ref{attractors}, the delay embeddings for the 3GP and DNLS current show clear tori in the case of Rabi  {regime} (\ref{attractors}a and \ref{attractors}d) and self-trapping dynamics (\ref{attractors}c and \ref{attractors}f). However, the chaotic regime explores a much larger region of the reconstructed phase space, while also displaying fractal-like structure (\ref{attractors}b and \ref{attractors}e). In contrast to the particle current, the local density appears spherical during the Rabi regime (see Fig.~\ref{attractors}g  and Fig.~\ref{attractors}i), while the chaotic regime, (see Fig.~\ref{attractors}h) is much more dense, and qualitatively different from its counterparts.

\par The correlations sums are displayed in Fig.~\ref{corrsum}, where we have selected the Theiler window $w=\sqrt{10^5}\approx33 T_r$ in order to avoid temporal correlations, for each of the attractors given in Fig.~\ref{attractors}. All show a region of linear scaling with $\epsilon$ on the log-log scale identified by the vertical dashed lines. When such a region is found, the size of $\epsilon$ must be within the bounds of the attractor, otherwise an incorrect dimension will be returned; e.g. for large the $\epsilon$ seen in Fig.~\ref{corrsum}(c) one would find $D=1$, which is clearly incorrect as the corresponding attractor (Fig.~\ref{attractors}(c)) can be seen to be dense on a 2D surface. For each measure and interaction strength tested that  {can be identified with a} linear scaling, we  {perform a fit} with a linear regression and extract the correlation dimension $D_{2}$, keeping those $D_{2}$ which have an $R^2\geq{}0.99$.  In Fig.~\ref{d2plat} (a)-(i) we see that a plateau is observed as $m$ is increased. This is indicative of the convergence of $D_{2}$. However,  {one finds} a 
slow increases of the correlation dimension for some time series after the plateau is  {reached}. This is typical in time-series of fine length with some amount of noise~\cite{theiler1990}. In order to account for this slow increase, we average over the plateaued region, including the error in the linear fits, which gives the final correlation dimension with some uncertainty seen in Tab.~\ref{d2ex}.  {From these three interaction strengths it is clear that higher dimensional attractors are observed for the DNLS local density (Fig.~\ref{attractors}(g)-(i)) when compared to the particle current.}

\begin{table}
\begin{tabular}{ c || c | c | c }
Model and & Rabi & Chaos & self-trapping  \\
 Measure & $g=0.03$ & $g=0.14$ & $g=0.34$ \\
 \hline 3GP Current & $2.06^{\pm0.02}$ & $3.27^{\pm0.13}$ & $2.02^{\pm0.02}$ \\
 DNLS Current & $2.07^{\pm0.02}$ & $2.97^{\pm0.05}$ & $2.30^{\pm0.02}$\\
DNLS Density $|\phi_3|^2$ & $2.59^{\pm0.01}$ & $3.93^{\pm0.04}$ & $3.05^{\pm0.05}$
\end{tabular}
\caption{\label{d2ex}\emph{Correlation Dimension.} Saturation value for the correlation dimension, $D_{2}$, of the attractors seen in Fig.~\ref{attractors}. The dimension for 3GP and DNLS particle  {currents} are in general agreement. Small deviations are expected, as the 3GP removes the short time dynamics due to driving. The local density of the DNLS is consistently larger than the corresponding total current, due to oscillations on the time-scale of the driving that are averaged out for the latter measure.}
\end{table}

\par
\begin{figure}
\begin{center}
\includegraphics[width=\linewidth]{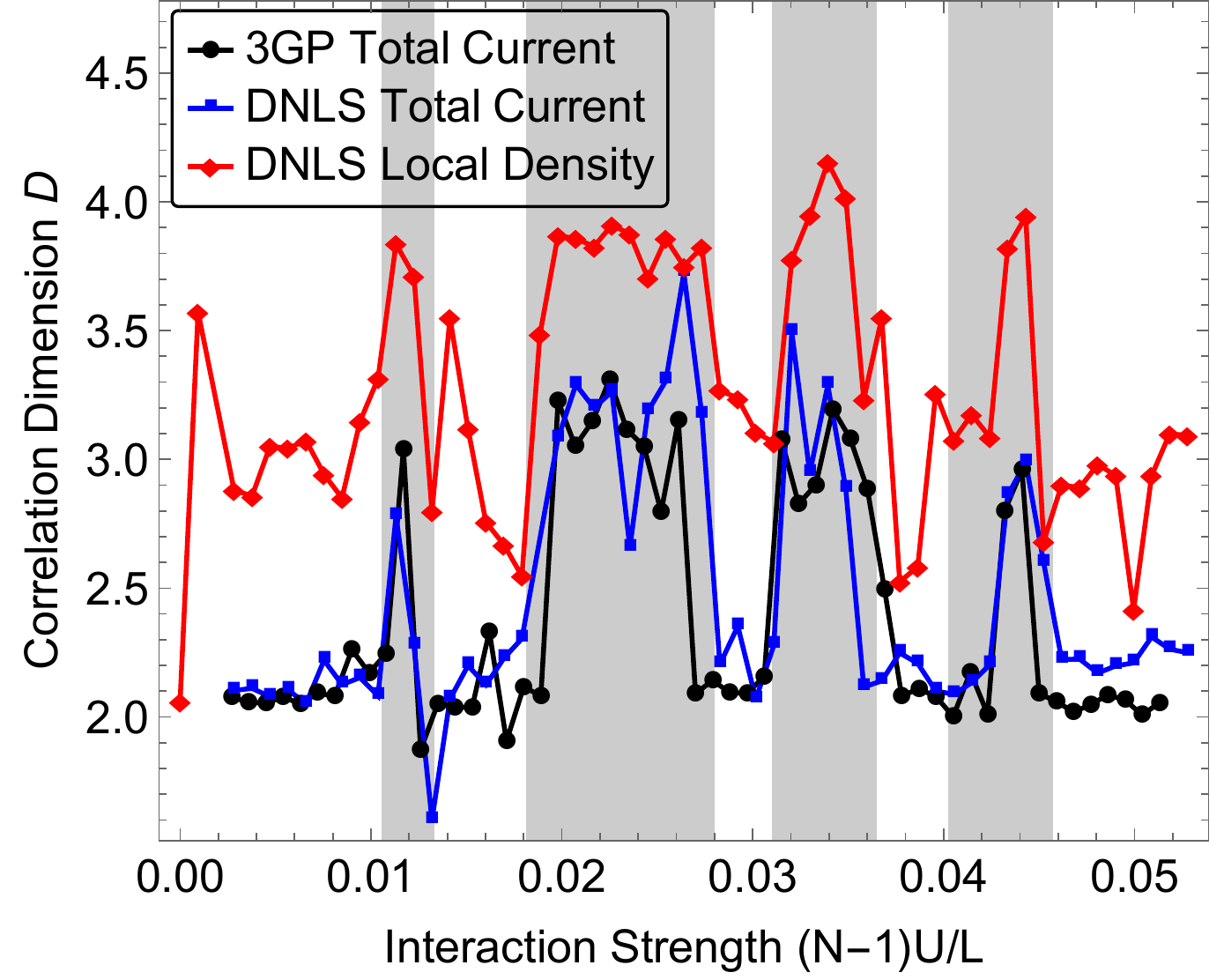}
\end{center}
\caption{\label{fractaldim}\emph{Mean-Field Fractal Dimension.} The correlation dimension, $D_{2}$, shows sharp increases for the regimes where chaos was identified by the ${0}-{1}$ test for chaos (vertical gray regions). The dimensions of the current in the 3GP and the DNLS follow the same trend with increasing interaction strength, further reaffirming the effectiveness of the three mode model. However, the local density in the DNLS is seen to be greater that the particle current regardless of interaction. This is due to the inclusion of rapid oscillations with the drive potential that get averaged out when the whole lattice is considered. Lines are meant only to guide the eye.}
\end{figure}

 Figure~\ref{fractaldim} gives the correlation dimension for the 3GP current, DNLS current, and DNLS local density for interaction strengths ranging fro $g=0$ to 0.34 with lines meant to guide the eye. It is clear that an increased correlation dimension is associated with chaotic dynamics as identified by the ${0}-{1}$ test for the gray shaded regions. We stress that the particle current in the 3GP and DNLS are in general agreement, with only slight variations. The local density in the DNLS is greater than the particle current by $\delta{D}\simeq{1}$ for each interaction strength. This implies that if one were to experimentally observe a system such as this, with two vastly different time scales, the reconstructed attractor dimension can vary with the layer of the system observed, i.e., global versus local. In our system, this can be explained by accounting for the fact that the particle current averages over the entire lattice in the DNLS, and that the fluctuations due to the driving are averaged out. Moreover, we note that the drive period $T$ and the Rabi period $T_\tx{R}$ cannot be related by a rational number. This implies that a time series which contains the shorter time scale must have an attractor that is one dimension larger for regular dynamics, which is clearly seen if one observes the weakly interacting regime at $g=0.02$.

\section{Characterizing the Approach to Mean-field}

\par
\begin{figure}
\begin{center}
\includegraphics[width=\linewidth]{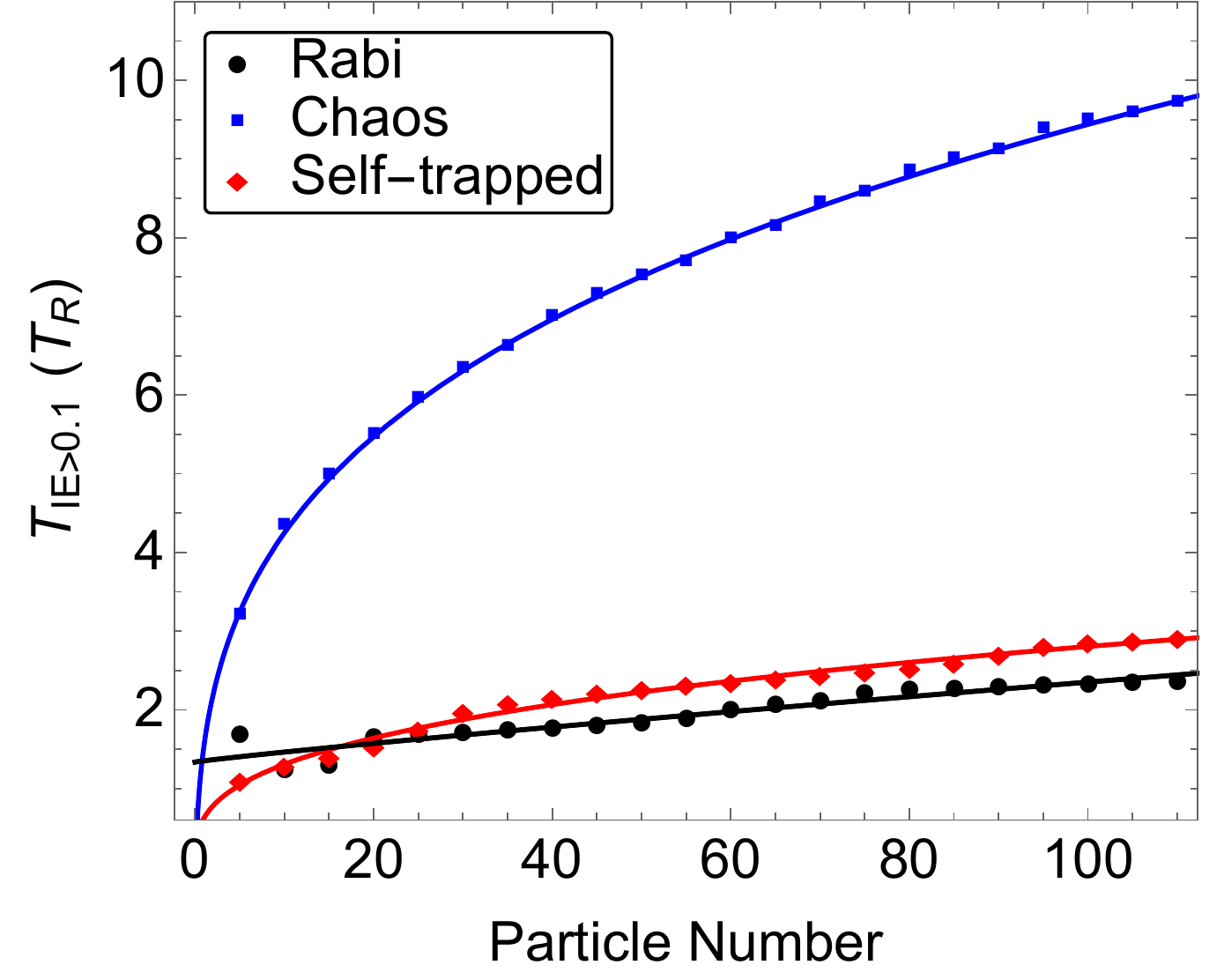}
\end{center}
\caption{\label{3ls-3gp-cur}\emph{Convergence to Mean-Field.} Integrated error of the normalized 3LS particle current compared to the 3GP particle current for $g=0.03$, 0.14, and 0.34. The  {considered interactions} give an overall measure for the convergence to mean-field in the Rabi, chaotic, and self-trapping regimes, respectively. Each regime has a polynomial approach to mean-field, with power laws of $b=0.28\pm0.01$, $0.34\pm0.02$, and $b=0.90\pm0.24$, respectively.}
\end{figure}

 With the mean-field attractors characterized, in this section we test the convergence of our many-body quantum ratchet to its mean-field limit. Many studies have shown the convergence of the Bose-Hubbard model to the DNLS by means of MPS methods~\cite{alcala2017}. However, as mentioned in our previous work~\cite{valdez2017}, the requirements of full local dimension and the vast difference in time scales renders the applicability of time evolution methods such as the Suzuki-Trotter expansion highly inefficient. This limitation, along with the fact that the 3LS has been well tested dynamically~\cite{heimsoth2013, valdez2017}, means that we can consider the 3LS as our testing ground for the convergence to mean-field.  {In Fig.~\ref{3ls-3gp-cur}, 
we plot the integrated error of the normalized 3LS particle current as compared to the 3GP particle current for various strengths $g$.} Here we have renormalized the 3LS current by the particle number, such that it matches the 3GP normalization.  {Figure ~\ref{3ls-3gp-cur} 
clearly shows that each regime has polynomial  {scaling} and can be  {fitted} with a function of the form $T_{\tx{IE}<0.1}=a\tx{ }N^b+c$.}  {The Rabi regime with the coupling strength $g=0.03$ gives $b=0.283\pm0.013$, while the chaotic regimes with the strength $g=0.14$ returns $b=0.342\pm0.066$, and finally the self-trapping regime with the coupling strength $g=0.34$ yields $b=0.90\pm0.236$.}

\par
\begin{figure}
\begin{center}
\includegraphics[width=\linewidth]{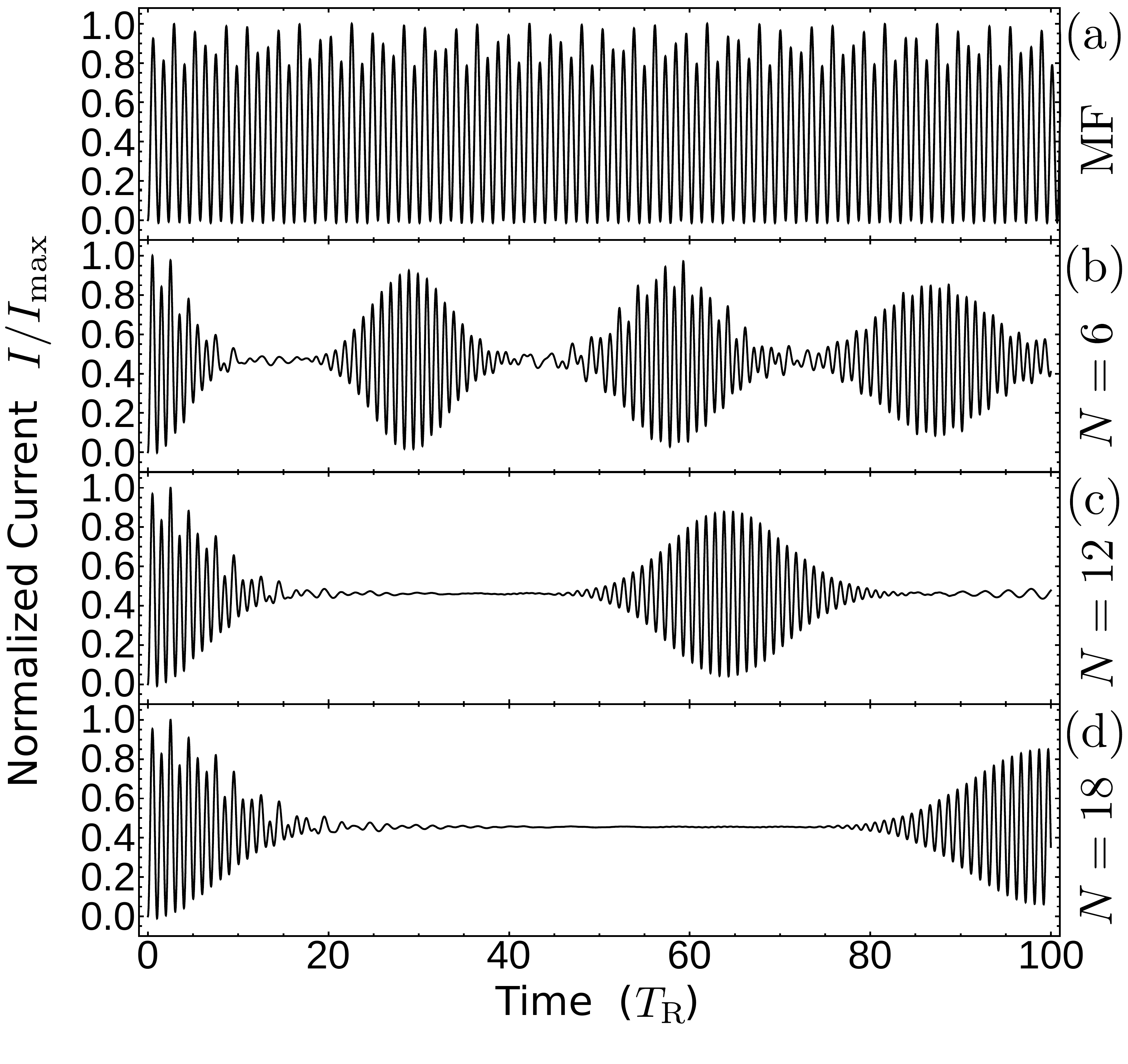}
\end{center}
\caption{\label{current}\emph{Dynamics Beyond Convergence Time.} (a)-(d) share the horizontal axis given by (d). (a) The normalized particle current, $I/I_\tx{max}$, for the mean-field 3GP. (b) The 3LS for particle number $N=6$ has an envelope that decays to a nearly constant value beyond the initial time it is converged to the mean-field result. This is then followed by a quantum revival of nearly Rabi-like oscillations, then the pattern repeats. (c)-(d) Normalized particle current $N=12$ and 16, respectively, show the same qualitative behavior as $N=6$, with envelopes and regions of near constant current increasing with $N$.}
\end{figure}
\begin{figure}
\begin{center}
\includegraphics[width=\linewidth]{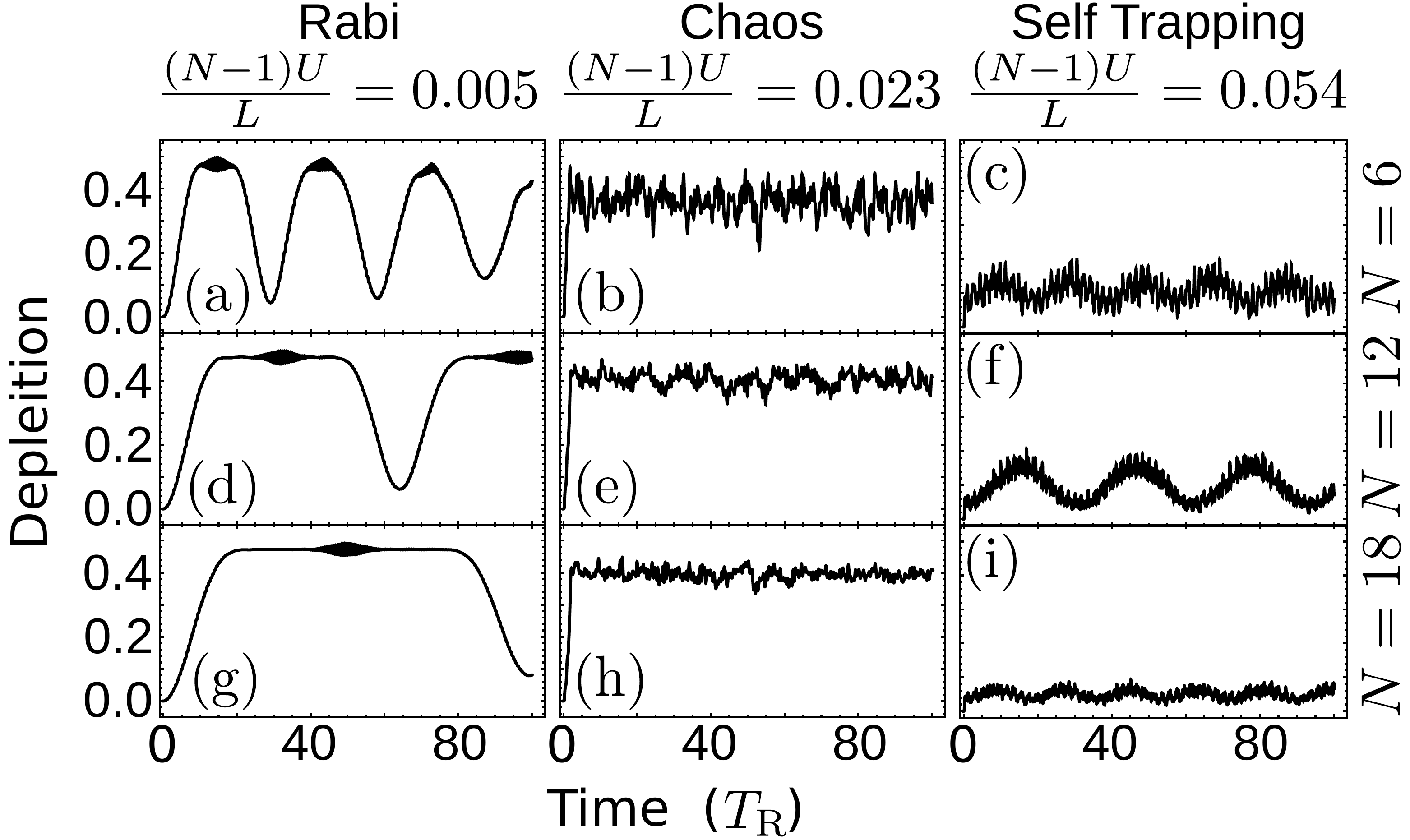}
\end{center}
\caption{\label{depletion}\emph{Condensate Depletion.} (a)-(c) Depletion of the original condensate for each dynamical regime in the 3LS.
 {The depletion is defined as $D=1-{\lambda_{1}}/{N}$, where $\lambda_{1}/N$ is the ratio of the largest eigenvalue of the single particle density matrix
$\langle{}\hat{a}_{i}\hat{a}_{j}{}\rangle$ to a total number of particles, $N$, that form  {the} BEC. Physically, depletion measures the portion of the BEC that remains in a single-particle mode. The large values of depletion throughout the dynamical regimes reveal
significant deviations from the mean-field description of the BEC. Thus, persistent
depletion indicates many-body dynamics of the condensate that cannot be captured within the mean-field approximation}.
(d)-(f) and (g)-(i) are for the same mean-field interaction strength with increased particle number, i.e., $NU_\tx{3LS}=\tx{const}$. The Rabi and self-trapping regimes show clear quantum revivals of a condensate, while the chaotic regime remains depleted over multiple macroscopic modes.}
\end{figure}

 {From the 3LS it is clear that the many-body dynamics approach the mean-field limit. However, the applicability of the sub-unity scaling approach is limited
for testing the convergence of the correlation dimension.  {Therefore}, we are led to explore the dynamics of the 3LS for time $t_{\text {in}}$ constrained from below by the  convergence time, i.e., $t_{\text {in}}>T_{\tx{IE}}$, while not exceeding the time necessary for a delay embedded reconstruction.} In Fig.~\ref{current}(a)-(d) we plot the normalized particle current ($I/I_\tx{max}$) in the Rabi regime (g=0.03) over 100$T_\tx{R}$ for the 3GP and the 3LS with $N=6$, 12, and 18 , respectively.  {For the $N=6$, we  {observe} beat patterns, where the current is approximated by the mean-field Rabi dynamics envelopes separated by a nearly constant particle current in between.} As the number of particles is increased, we see  {an} elongation of the envelope of the oscillations, with further elongated regions of nearly constant current. Extending the time range for the larger system sizes reveals revivals of current similar to those seen for $N=6$. The onset of nearly constant current is indicative of a change in the underlying structure of the many-body dynamics. Indeed, if we plot the depletion (Fig.~\ref{depletion}(a)), defined as $D=1- \lambda_1/N$, where $\lambda_1$ is the largest eigenvalue of the single particle density matrix $\langle \hat{a}^\dag_\mu \hat{a}_\nu \rangle$, for this same time range and particle numbers, we see high depletion for the constant current ranges. The ranges where oscillating current is observed can be seen to have the depletion trend with the envelope. It is clear that the coherent dynamics of the quantum ratchet are short lived, indicating that reconstructing the systems mean-field attractor from quantum many-body dynamics would require the initial envelope to range the entire 1000 $T_\tx{R}$ timescale.

\par For the increased interaction strength into the chaotic regime ($g=0.14$), we again observe the onset of high depletion, see Fig.~\ref{depletion}(b) with $N=6$, 12, and 18, respectively. In contrast to the Rabi regime, there is no decay of the depletion back to near zero for the chaotic dynamics within 100$T_\tx{R}$. For the self-trapping regime, the depletion increases, but not to the magnitude of  {Rabi or chaotic dynamics}, with a maximum of approximately 0.2.  {Similar to our} previous finding, the depletion of the condensate for these regimes makes the reconstruction of delay embedded attractors not possible for accessible system sizes and time scales, due to insufficient attractor sampling.

\par
\begin{figure}
\begin{center}
\includegraphics[width=\linewidth]{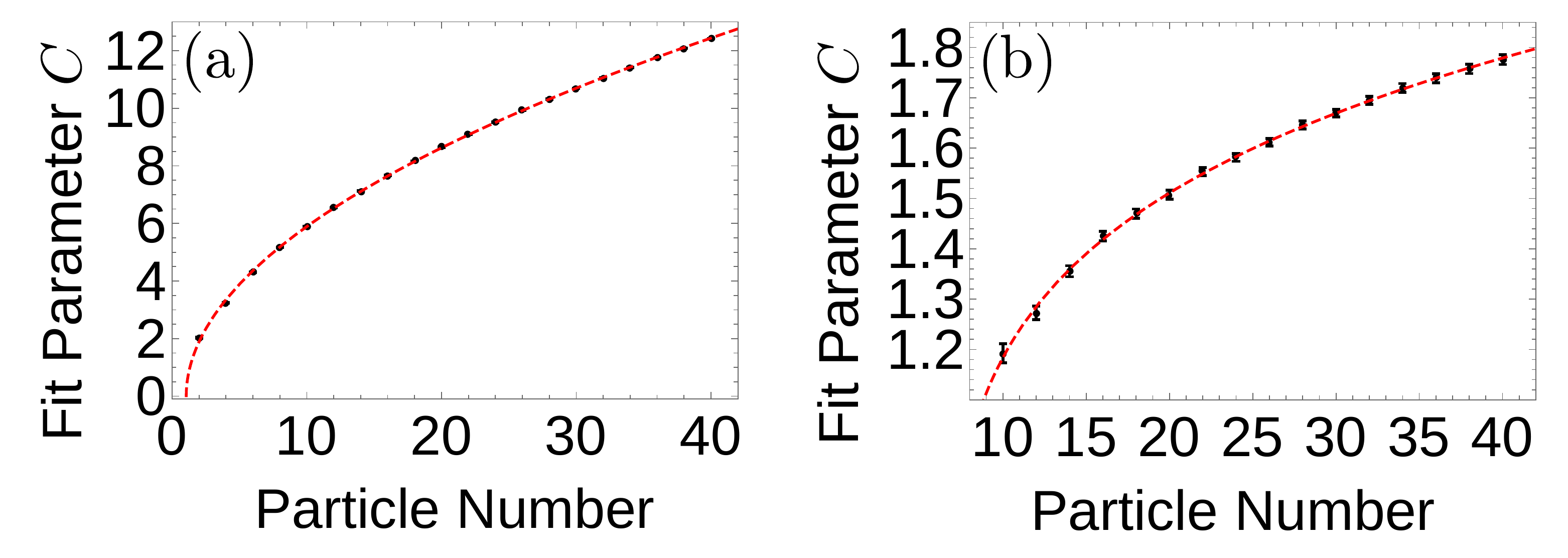}
\end{center}
\caption{\label{deptime}\emph{Onset of Depletion} The black points are data for $C$ and the red curves are the fitting functions. Figures (a) and (b) show the onset of depletion in the Rabi and chaotic regimes, respectively, as measured by the turning point of the $\tanh$ fit. Both scale polynomially with particle number $N$. For Rabi dynamics the onset of depletion scales with $N^{0.51\pm0.004}$,  {while for chaotic dynamics it scales as} $N^{0.18\pm0.004}$.}
\end{figure}

\begin{figure}
\begin{center}
\includegraphics[width=\linewidth]{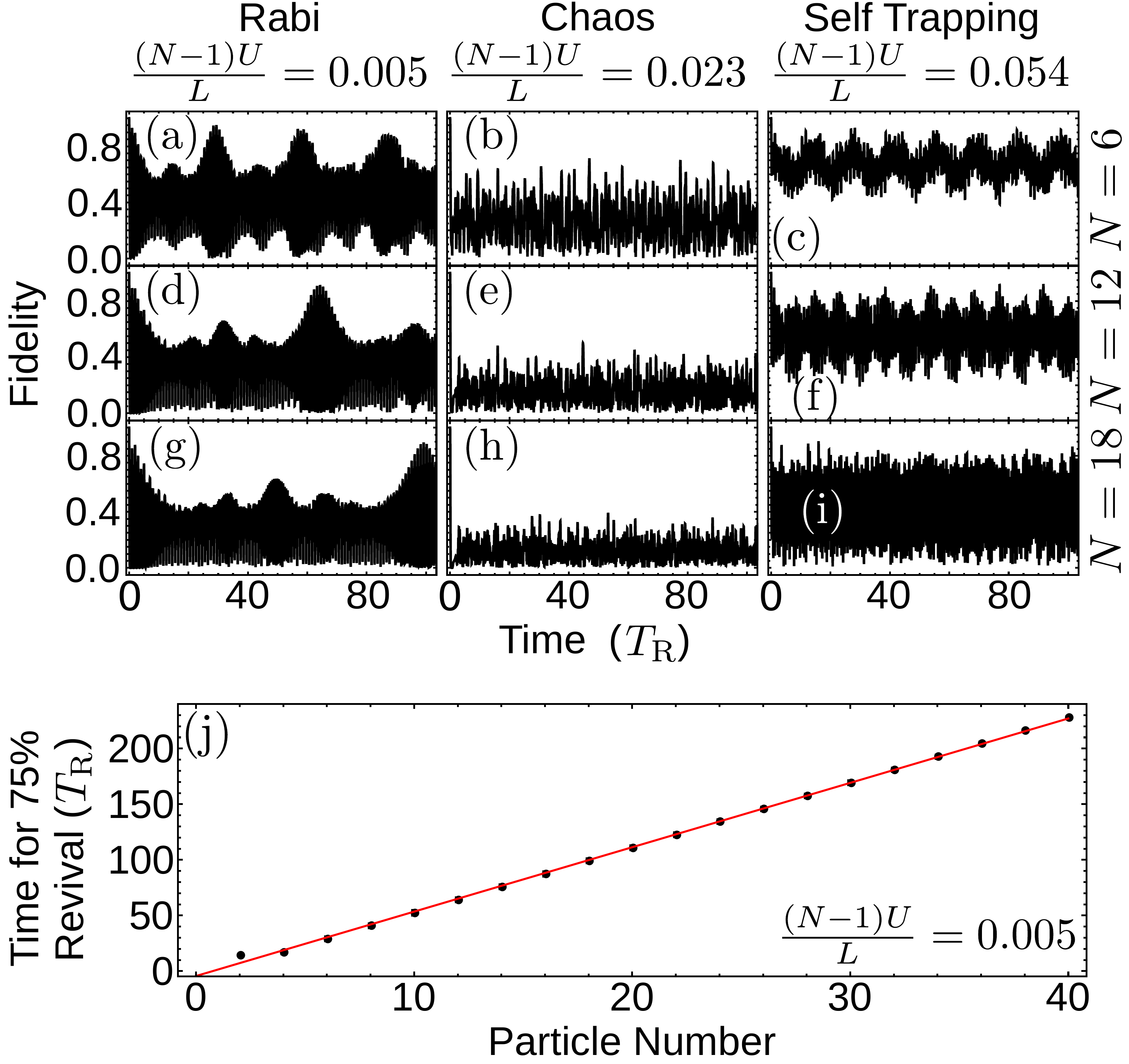}
\end{center}
\caption{\label{fidelity}\emph{Fidelity.} (a)-(c) Overlap of the system with its initial state, $|\langle \psi(0)|\psi(t)\rangle|$, for Rabi, chaotic, and self-trapping dynamics, respectively. (c)-(f) and (g)-(i) give the same measure for increased system size, with columns corresponding to regimes. The Rabi and self-trapping regimes have large overlaps with the initial condensate, with the revivals in the  {Rabi regime} corresponding to the decrease in depletion, see Fig.~\ref{depletion}. Chaotic dynamics  {have little overlap with the initial condensate, with the fidelity decreasing as the particle number is increased.} (j) Mean time for $|\langle \psi(0)|\psi(t)\rangle|\geq0.75$ in the Rabi regime increases linearly in particle number, following $T_\tx{revive}=5.79 N - 4.41$.}
\end{figure}

 The initial onset of depletion for the Rabi and chaotic regimes can be fitted with  {the} function $D(x)=A \tanh[B(x+C)]+D$. From these fits we extract value for $C$, which is the turning point of the $\tanh$ function, and for increasing system sizes gives a measure of the time for which a coherent condensate is present. On a log-log scale it is clear that $C$ scales polynomially with particle number $N$. Using the fitting function $C = \alpha (N+\beta)^\delta$, see Fig.~\ref{deptime}, we find that the onset of depletion gives $\beta = 0.508\pm0.004$ for Rabi dynamics and $\beta = 0.179\pm0.004$. We note that the system sizes $N=2$, 4 , 6, and 8 have not been used for the chaotic regime, and that the self-trapping regime is not included since it could not be properly
fitted for $D(t)$. The trends of depletion after its initial onset seen in Fig.~\ref{depletion} indicate that the condensate has revivals for the Rabi and self-trapping regimes, while the chaotic regime does not. If we plot the fidelity defined as $|\langle \psi(t=0)|\psi(t)\rangle|$, seen in Fig~\ref{fidelity}a-i, we confirm that  {the} Rabi and self-trapping dynamics have   {a} large overlap with the initial condensate, while chaotic dynamics present no revivals. For the first case, a clear envelope of revival is seen. The other two simply oscillate rapidly with no discernible pattern. In the case with clear trending, we extract the mean time for the first revival of $|\langle \psi(0)|\psi(t)\rangle|\geq0.75$, and give it as a function of $N$ in Fig~\ref{fidelity}j, which has a linear scaling with the particle number, namely $T_\tx{R}=-4.41 + 5.79N$. We note the average was taken due to the highly oscillatory nature of the measure within the broad revival peak.

\section{Conclusions and Outlook}

In conclusion, we explored a driven Bose-Einstein condensate quantum ratchet and identified four distinct pockets of chaos on the interval of interaction strength ranging from $g=0$ to $g=0.28$. These regions are identified for the same interaction strengths using
 the particle current in an effective three level mean-field model and the discrete nonlinear Schr\"{o}dinger equation,  {as well as} the local density of the discrete nonlinear Schr\"{o}dinger equation. Our study of the particle current reveals that the three level mean-field model and discrete nonlinear Schr\"{o}dinger equation have qualitatively similar trends in correlation dimension for the interaction strengths considered. However, the local density in the discrete nonlinear Schr\"{o}dinger equation results in a dimension that is consistently higher. We characterized the approach of the many-body three level system to the three level mean-field model via integrated error, finding a sub-unity power law scaling with particle number for Rabi, chaotic, and self-trapping dynamics. Specifically, $N^ {0.28\pm0.013}$, $N^{0.34\pm0.066}$ and $N^{0.90\pm0.236}$ are observed for the Rabi,  chaos, and self-trapping regimes, respectively. We also characterizes the dynamics of each of the mentioned dynamical regimes beyond the convergence time, where the deviation from the mean-field value occurs due to
the onset of condensate depletion. In the Rabi and chaotic cases, the depletion has a $\tanh$-like onset, the turning points of which scale as $N^{0.51\pm0.004}$ and $N^{0.18\pm0.004}$, respectively. The decreases in depletion after its initial onset, was found to match the revival of over $75\%$ of the original condensate in the Rabi regime, with the revival time scaling linearly in $N$.  { In contrast,  {a} quantum ratchet with the coupling strength corresponding to the chaotic regime does not exhibit quantum revivals. Although the time scales necessary to reliably reconstruct the attractor of our quantum many-body system are currently out of reach for the current Bose-Einstein condensate-based experiments, the conclusions obtained in the paper will be important for the future studies of driven many-body systems. In fact, local oscillations and fluctuations that may be averaged out in macroscopic measures, e.g., current, can present higher dimensional attractors. Therefore, fractal structures in the dynamics of many-body systems depend on the observable, which thus establishes the notion of
layered chaos in these systems.}

 {The performed study of a driven quantum ratchet opens up new pathways in the exploration and characterization of emergent phenomena in driven many-body systems. The prime example is the identification of Anderson and many-body localization in driven interacting and non-interacting quantum many-body systems by means of many-body measures, including binary identification of chaotic dynamics, calculation of the depletion, and evaluation of the correlation dimension. The other important research pathway corresponds to the identification of the characteristic time scale at which mean field description fails to describe quantum many-body system. Specifically, the characteristic scaling of the Ehrenfest time, $\tau_{E}$, which can be expressed in terms of the Lyapunov exponent $\gamma$ and particle number $N$ as $\tau_{E}\simeq{}\log{(N)}/\gamma$, and which is conventionally used in the estimate of the breakdown of mean field description, was found to have large deviation in the special case of our system.  Therefore, the open question is whether the Ehrenfest time scales logarithmically or polynomially with the number of particles that form a quantum many-body interacting system. The other critical question is the dependence of the rate of convergence of the many-body dynamics to its mean-field limit on the amount of entanglement present in the system. Finally, an important question is related to the
rate of the convergence to the semiclassical strange attractors as a function of dissipation
exhibited by the system.}

\begin{acknowledgments}
 {
 {This material is based upon work supported by} the U.S. National Science Foundation under Grant Nos.  {} OAC-1740130,  {} CCF-1839232,  {and} PHY-1806372,  {by} the U.S. Air Force Office of Scientific Research Grant number FA9550-14-1-0287,
and  {by} the U.K. Engineering and Physical Sciences Research Council (EPSRC) through
the ``Quantum  Science  with  Ultracold  Molecules" Programme (Grant No. EP/P01058X/1). This work has been supported by Spain's MICINN through Grant Nos. FIS2013-41716-P and FIS2017-84368-P. One of us (FS) would like to acknowledge the support of the Real Colegio Complutense at Harvard and the Harvard-MIT Center for Ultracold Atoms, where part of this work was done. The calculations were executed on the high performance computing cluster maintained by the Golden Energy Computing Organization at the Colorado School of Mines.}
\end{acknowledgments}


%

\end{document}